 \documentclass[12pt,preprint]{aastex}

\def\dsm{$\mathrm{M}_\odot$}
\def\dsr{$\mathrm{R}_\odot$}

\shorttitle{contributions of interacting binary stars to DMSTOs and
dual RC} \shortauthors{Yang et al.}

\begin{document}


\title{THE CONTRIBUTIONS OF INTERACTIVE BINARY STARS TO DOUBLE MAIN SEQUENCE
TURN-OFFS AND DUAL RED CLUMP OF INTERMEDIATE-AGE STAR CLUSTERS}

\author{Wuming Yang\altaffilmark{1,2}, Xiangcun Meng\altaffilmark{2},
Shaolan Bi\altaffilmark{1}, Zhijia Tian\altaffilmark{1}, Tanda
Li\altaffilmark{1}, Kang Liu\altaffilmark{1}}

\altaffiltext{1}{Department of Astronomy, Beijing Normal University, Beijing
100875, China; yangwuming@ynao.ac.cn, woomyang@gmail.com}
\altaffiltext{2}{School of Physics and Chemistry, Henan Polytechnic University,
Jiaozuo 454000, Henan, China}

\begin{abstract}
Double or extended main-sequence turn-offs (DMSTOs) and dual red
clump (RC) were observed in intermediate-age clusters, such as in
NGC 1846 and 419. the DMSTOs are interpreted as that the cluster has
two distinct stellar populations with differences in age of about
200-300 Myr but with the same metallicity. The dual RC is
interpreted as a result of a prolonged star formation. Using a
stellar population-synthesis method, we calculated the evolutions of
binary-star stellar population (BSP). We found that binary
interactions and merging can reproduce the dual RC in the
color-magnitude diagrams of an intermediate-age cluster, whereas in
actuality only a single population exists. Moreover, the binary
interactions can lead to an extended MSTO rather than DMSTOs.
However, the rest of main sequence, subgiant branch and first giant
branch are hardly spread by the binary interactions. Part of the
observed dual RC and extended MSTO may be the results of binary
interactions and merger.
\end{abstract}


\section{Introduction}

It has always been accepted that a star cluster (SC) is composed of
stars belonging to a single, simple stellar population with a
uniform age and chemical composition. In recent years, however, due
to the increase in spatial resolution and photometric accuracy, it
has been discovered that some SCs have unusual structures in their
observed color-magnitude diagrams (CMDs). For example, NGC 2173 in
the Large Magellanic Cloud (LMC) has an unusually large spread in
color about its main-sequence turn-off (MSTO) \citep{bert03}; Omega
Centauri exhibits a main sequence (MS) bifurcation \citep{piot05};
NGC 2808 possesses a triple MS split \citep{piot07}; and other
clusters also contain mutiple stellar populations \citep{milo09,
milo10}. More interesting discoveries are that the intermediate-age
massive clusters in the LMC, such as NGC 1846, 1806, and 1783,
possess double main-sequence turn-offs (DMSTOs) on their CMDs
\citep{mack07, mack08, milo09}. Moreover, \cite{goud09} found that
the spread of the MSTO is fairly continuous rather than strongly
bimodal. \cite{mack08} and \cite{goud09} argued that these clusters
do not possess a significant line-of-sigth depth or internal
dispersion in [Fe/H], or suffer from significant differential
extinction. \cite{mucc08} showed that spreads in [Fe/H] in SCs like
NGC 1783 are very small. The apparent homogeneity in [Fe/H] of all
stars in NGC 1846 indicates that capture of pre-existing field stars
during cluster formation seems to hard to explain the DMSTOs
\citep{mack08, goud09}. The DMSTOs are interpreted as that the
clusters have two distinct stellar populations with differences in
age of $\sim$ 200-300 Myr but with similar metal abundance
\citep{mack07, mack08}. Moreover, \cite{glat08} found multiple MSTOs
(MMSTOs) in NGC 419 in the Small Magellanic Cloud, and \cite{gira09}
found two distinct red clumps (RC) in this cluster. These
observational results are challenging the traditional picture.

In order to understand how a cluster can contain multiple
populations with different ages, many scenarios were proposed.
\cite{mack07} put forward the scenario of merger of two (or more)
SCs. However, \cite{goud09} pointed out that this scenario seems
unlikely. \cite{bekk09} simulated the merger of a SC with a giant
molecular cloud (GMC). They found that this merger can produce a
second-generation of stars that are required to explain the DMSTOs.
However, in order to obtain the differences in age of $\sim$ 200 -
300 Myr, this scenario require rather strongly constrained ranges of
GMC parameters such as their spatial distribution, mass function,
and chemical composition \citep{goud09, bast09}. \cite{derc08} and
\cite{goud09} put forward the scenario of formation of a second
generation of stars from the ejecta of first generation asymptotic
giant branch stars. In this scenario, the young second generation is
usually less than the first one \citep{bast09}, which is not
consistent with observations \citep{milo09}. Recently, using two
scaling relations to mimic the effects of rotation, \cite{bast09}
considered the effect of rotation on stellar evolutions. They found
that stellar rotation in stars with masses between 1.2 and 1.7
\dsm{} can mimic the effect of a double population when rotation
rates reach 20-50 per cent of the critical rotation, whereas in
actuality only a single population exists. However, \cite{rube10}
and \cite{gira11} argued that rotational effect could not explain
the presence of MMSTOs. They proposed a prolonged star formation
(PSF) to explain the MMSTOs and the dual RC. In \cite{rube10}
models, in order to reproduce the dual RC of NGC 419, a PSF of about
700 Myr is required. The dual RC was also found in NGC 1751
\citep{rube11}. In order to explain this dual RC, a PSF of about 460
Myr is required \citep{rube11}. This age spread is twice longer than
$\sim$ 200 Myr estimated by \cite{milo09} using the method of
isochrone fitting. Furthermore, recently, \cite{kell11} argued that
the isolation of extended MSTO clusters to intermediate ages could
be the consequence of observational selection effects.

In addition, binary stars of LMC clusters are unresolved even with
Hubble Space Telescope \citep{milo09}. The binary system should
appear as a single point-source object. Thus the unresolved binary
systems may have contributions to the spread of MSTO. However,
several investigators \citep{mack08, goud09, milo09} had found that
unresolved binaries cannot alone reproduce the peculiar MSTO
structures. Nevertheless, after a star in a binary system accretes
mass from its companion, it will become younger apparently. Hence,
interactive binaries may have contributions to the spread of MSTO.
\cite{goud09} found that the younger populations in clusters are
more centrally concentrated than the older populations. Thus the
interactive binaries may be important in interpreting the observed
CMDs. In this paper, using the Hurley rapid single and binary
evolution codes \citep{hurl00,hurl02} and a stellar
population-synthesis method, we investigated the contributions of
interactive binaries to the DMSTOs and dual RC.

The paper is organized as follow. We show our stellar
population-synthesis method in section 2. We present the results in
section 3 and discuss and summarize them in section 4.

\section{Stellar population synthesis}

To investigate DMSTOs and dual RC of SCs, we calculated a
binary-star stellar population (BSP). Stellar samples are generated
by Monte Carlo simulations. The basic assumptions for the
simulations are as follows. (i) The lognormal initial mass function
(IMF) of \cite{chab01} is adopted. (ii) We generate the mass of the
primary, $M_{1}$, according to the IMF. The ratio ($q$) of the mass
of the secondary to that of the primary is assumed to be a uniform
distribution within 0-1 for simplicity. The mass of the secondary
star is then determined by $qM_{1}$. (iii) We assume that all stars
are members of binary systems and that the distribution of
separations ($a$) is constant in $\log a$ for wide binaries and
falls off smoothly at close separation:
\begin{equation}
an(a)=\left\{
 \begin{array}{lc}
 \alpha_{\rm sep}(a/a_{\rm 0})^{\rm m} & a\leq a_{\rm 0};\\
\alpha_{\rm sep}, & a_{\rm 0}<a<a_{\rm 1},\\
\end{array}\right.
\end{equation}
where $\alpha_{\rm sep}\approx0.070$, $a_{\rm 0}=10R_{\odot}$,
$a_{\rm 1}=5.75\times 10^{\rm 6}R_{\odot}=0.13{\rm pc}$ and
$m\approx1.2$. This distribution implies that the numbers of wide
binary system per logarithmic interval are equal, and that
approximately 50\% of the stellar systems are binary systems with
orbital periods less than 100 yr \citep{han95}. \textbf{However,
binaries account for typically $\sim$30-40\% of all stars in the
clusters of LMC, such as in NGC 1818 and 1806 \citep{elson98,
milo09}.} (iv) The eccentricity (e) of each binary system is assumed
to be a uniform distribution within 0-1. With these assumptions, we
calculated the evolutions of 5$\times 10^{4}$ binaries with $M_{1}$
between 0.8 and 5.0 \dsm{}.

The metal abundance Z of evolutionary models was converted firstly
into [Fe/H]. Then the theoretical properties ([Fe/H], T$_{eff}$,
$\log g$, $\log L$) have been transformed into colors and magnitudes
using the color transformation tables of \cite{leje98}. The binaries
with $a \leq 10^6$ \dsr{} were treated as unresolved ones, while
others were treated as resolved ones when we computed their colors
and magnitudes.

\section{Calculation results}

\begin{figure}
\includegraphics[angle=-90, scale=0.8]{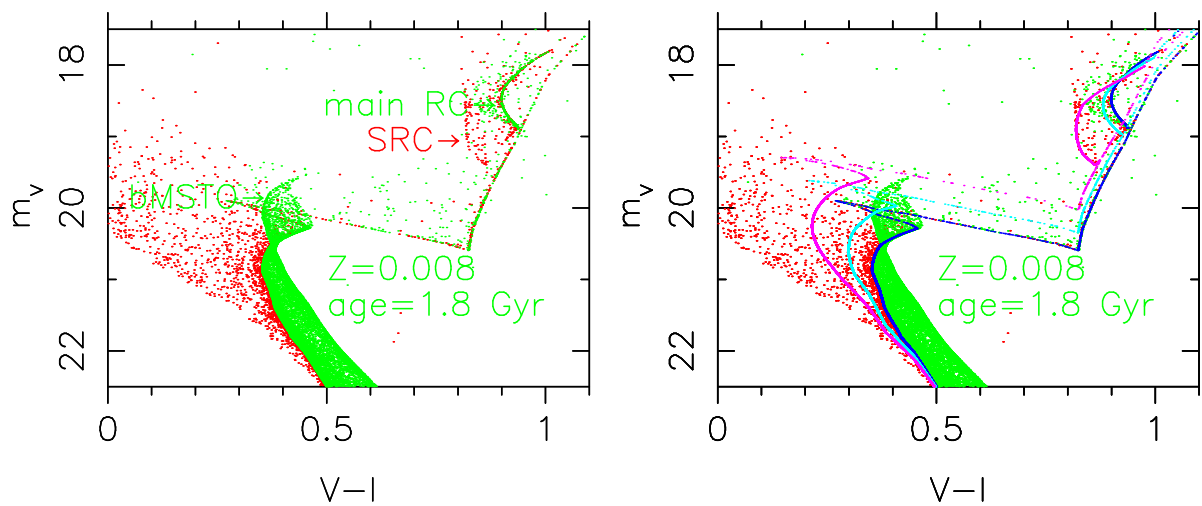}
\caption{The CMDs of simulated BSP. Green shows the binaries almost
without mass transfer, while red indicates interactive binaries.
A distance modulus of 18.45 is adopted. The isochrones of SSP with
the same Z but with different ages are overplotted on the CMD in
the right panel (blue: 1.8, cyan: 1.5 and magenta: 1.2 Gyr). \label{fig1}}
\end{figure}

Figure \ref{fig1} shows the CMDs of the simulated BSP with Z = 0.008
and age = 1.8 Gyr. The interactive binaries are shown in red, while
non-interactive binaries are shown in green. The population of the
brighter MSTO (bMSTO) is mainly from the unresolved binaries with
$q>$ 0.7 and no interactions. The isochrones of single-star stellar
populations (SSP) with the same Z but different ages are overplotted
on the CMD in the right panel of Fig. \ref{fig1}. The stars that
experienced binary interactions clearly deviate from the isochrone
of non-interactive binaries in the region around MSTO, which leads
to a large spread in color. The interactive binaries clearly produce
an extended MSTO rather than bimodal MSTO. For main sequence
\textbf{binary} stars with $\mathrm{m_{V}} <$ 22.5 mag,
\textbf{about 10\% of the binary stars lie between the 1.8 and 1.5
Gyr isochrones in our models, not including the binaries on the 1.8
Gyr isochrone}. However, interactive binaries hardly affect the rest
of MS, subgiant branch and first giant branch (FGB). Moreover,
interactive binaries result in the appearance of a secondary RC
(SRC) at about 0.5 mag below the main RC. The SRC is slightly bluer
than the ridgeline of FGB stars. The contribution of non-interactive
binaries to the SRC is negligible. The right panel of Fig.
\ref{fig1} shows that the ridge of the SRC is almost coincident with
the isochrone of RC stars of the SSP with age = 1.2 Gyr. About 12\%
of core-helium burning (CHeB) \textbf{binary} stars are located
between the 1.8 and 1.5 Gyr isochrones. However, about 30\% of CHeB
\textbf{binary} stars are located between the main RC and 1.2 Gyr
isochrone, not including \textbf{the binary} stars on the main RC
but including those on 1.2 Gyr isochrone. The mass of the SRC stars
is mainly located between 1.85 and 2.1 \dsm{}, while that of the
main RC stars is less than 1.7 \dsm{}.

\begin{figure}
\includegraphics[angle=-90, scale=0.8]{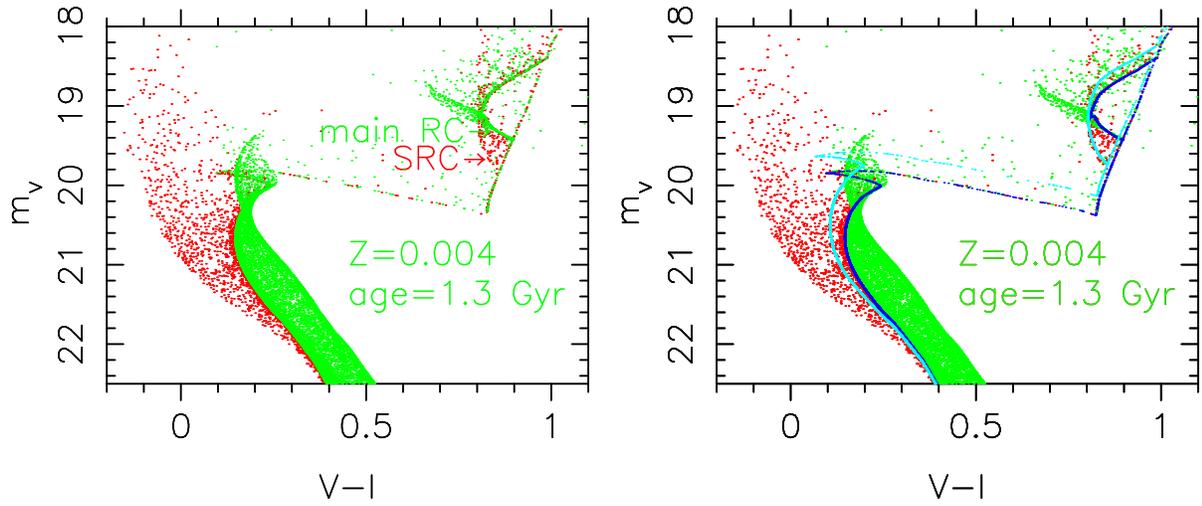}
\caption{Same as Fig. \ref{fig1}. A distance modulus of 18.84 is adopted.
The isochrones of SSP with the same Z but different ages are overplotted
on the CMD in the right panel (blue: 1.3 and cyan: 1.1 Gyr).
\label{fig2}}
\end{figure}

The dual RC was discovered in NGC 752, 7789 and 419 \citep{gira00,
gira09} and NGC 1751 \citep{rube11}. Cluster NGC 419 has an age of
about 1.35 Gyr \citep{gira09}. We computed a cluster with age = 1.3
Gyr and Z = 0.004. Its CMDs are shown in Fig. \ref{fig2}. Just as
above results, the interactive binaries lead to a large spread in
color in the region around MSTO, but they do not reproduce DMSTOs.
However, the interactive binaries lead to the presence of a distinct
SRC. The ridge of the SRC is almost coincident with the isochrone of
RC stars of the SSP with age = 1.1 Gyr. This implies that the SRC
produced by binary interactions has an apparent age of about 1.1 Gyr
in our models, whereas actually it has an age of 1.3 Gyr.
\textbf{For this dual RC, about 15\% of binary stars} are located
between the 1.3 and 1.1 Gyr isochrones, including the
\textbf{binaries} on the 1.1 Gyr isochrone but not including those
on 1.3 Gyr isochrone. The magnitude extension between main RC and
SRC is about 0.35 mag. The mass of the SRC stars is mainly located
between 1.85 and 2.05 \dsm{}, while that of the main RC stars is
mainly located between 1.78 and 1.81 \dsm{}. For Z = 0.004, in stars
more massive than about 1.88 \dsm{} (this value is about 2.0 \dsm{}
for Z = 0.02), helium-burning temperatures are reached at the center
before electrons become degenerate there. For these stars, when they
are on zero-age horizontal branch (ZAHB), their magnitudes decrease
with decreasing mass. However, for the stars with $M <$ 1.88 \dsm{},
electrons in the hydrogen-exhausted core are highly degenerate
before helium ignition occurs. When they are on ZAHB, their
magnitudes increase rapidly with decreasing mass except for stars
with $M <$ 1.5 \dsm{}. The magnitudes of ZAHB stars with $M <$ 1.5
\dsm{} increase very slowly with decreasing mass. For Z = 0.004, the
evolutionary tracks of RC stars with mass between about 1.85 and 2.0
\dsm{} are very near each other in CMDs. Thus the interactive
binaries with mass between about 1.85 and 2.0 \dsm{} can gather to
form a SRC. Our calculations show that the SRC caused by binary
interactions appears mainly in clusters with 1.2 Gyr $<$ age $<$ 3.0
Gyr.

\section{Discussion and Conclusions}

\cite{milo09} showed that the fraction of younger (brighter)
population is about 75\% in the case of NGC 1846. In our models, the
`younger' population are mainly from the merged binary systems and
interactive binaries with mass transfer. The merging and mass
transfer in binary systems can be affected by the distributions of
separation ($a$), eccentricity ($e$) and mass-ratio ($q$) of
systems. We computed the BSP by using the distributions $n(\log a)$
= constant \citep{hurl02}, $n(e) = 2e$ and $n(q) = 2q$ and the IMF
of \cite{salp55}. However, the fraction of the interactive binaries
are not obviously enhanced. We also computed the BSP using the
distribution $n(\log a)$ = constant and Gaussian distributions for
$e$ and $q$. The mean value and standard deviation of the Gaussian
distributions is 0.5 and 0.13 for $e$, 0.6 and 0.1 for $q$,
respectively. The values of the standard deviation are chosen in
order to make the values of $e$ and $q$ are located between 0 and 1.
In this simulation, \textbf{about 20\% of binary stars lie between
the 1.8 and 1.5 Gyr isochrones for the cluster with Z = 0.008 and
age = 1.8 Gyr; while for the RC of the cluster with Z = 0.004 and
age = 1.3 Gyr, about 30\% of binary stars lie between the 1.3 and
1.1 Gyr isochrones.}

The mass of turn-off stars is about 1.46 \dsm{} for the cluster with
Z = 0.008 and age = 1.8 Gyr in our models. The evolutions of a wide
binary system without mass transfer are similar to those of single
stars with the same mass. However, the evolution of stars whose mass
increased via mass accretion or merging is always slower than that
of single stars with the same mass. For example, a mass-accreted
star with $M$ = 1.55 \dsm{}, which was reached by accretion, just
arrives at around MSTO at the age of 1.8 Gyr. However, a single star
with an initial mass of 1.55 \dsm{} has left the MSTO at the same
age. Thus the mass-accreted stars clearly deviate from the ridge of
the isochrone in the region around MSTO and have an apparent younger
age. The more the accreted mass, the more the deviation. Thus the
spread of MSTO of BSP is continuous rather than bimodal. However,
for the mass-accreted stars with $M <$ 1.25 \dsm{}, an increase in
mass leads to a shift of their position in CMDs almost along MS.
Therefore, the most of MS are hardly spread by binary interactions.

The SRC is almost made of the mass-accreted or merged stars whose
mass is slightly more massive than the critical mass that is just
enough to avoid electron-degeneracy occurring in stellar H-exhausted
cores in our models. The stars with the critical mass usually evolve
into RC at the age of about 1.1-1.2 Gyr (For the stars with $Z \geq
0.04$, this age can increase to about 1.3 Gyr). Hence, the SRC
caused by interactive binaries should have an apparent age of about
1.1-1.2 Gyr, whereas actually it has an age as old as that of main
RC. The apparent age of about 1.1-1.2 Gyr is consistent with the
estimates of \cite{rube11, rube10} for NGC 1751 and 419. The main RC
stars passed through electron degeneracy. The magnitude of these
stars increases with decreasing mass. The older the age of clusters,
the lower the mass of main RC stars. Thus, the magnitude extension
between main RC and SRC increases with age until the mass of the
main RC is less than about 1.5 \dsm{}. Because the magnitude of ZAHB
stars with $M <$ 1.5 \dsm{} increases very slowly with decreasing
mass, there is an upper limit of magnitude extension of about 0.8
mag between main RC and SRC in our models. Moreover, with increasing
age, the SRC becomes slightly bluer than FGB. For the clusters with
age $<$ 1.2 Gyr, the mass of all RC stars is almost larger than the
critical mass. Therefore, the SRC caused by interactive binaries
should not appear in these clusters. In addition, in old clusters,
the initial mass of RC stars is low. If the mass is not enough to be
increased to above the critical mass by accretion or merging, the
SRC should not be produced by binary interactions in the old
clusters. Moreover, if the fraction of binaries is too low, the SRC
also could not be produced. \cite{gira09} argued that the MS+red
clump binaries cannot mimic the dual RC in NGC 419. In our models,
the SRC stars are from the merged binaries and binary systems with
mass transfer.

In this paper, we showed that binary interactions such as mass
transfer and binary merging can produce an extended MSTO and dual RC
in CMDs of intermediate-age clusters, whereas in actuality only a
single population exists. Despite these, the rest of MS, subgiant
branch and FGB are not clearly spread by the binary interactions.
Interactive binaries can lead to an extension of MSTO rather than
DMSTOs. For a cluster with Z = 0.008 and age = 1.8 Gyr, its
isochrone can be spread down to $\sim$ 1.5 Gyr by interactive
binaries. However, \textbf{only about 10\% of binary stars lie
between the 1.8 and 1.5 Gyr isochrones} in our models. Moreover, the
SRC that is caused by interactive binaries should have an apparent
age of about 1.1-1.2 Gyr. Although binary interactions cannot lead
to the bimodal MSTO, the SRC in NGC 419 and extended MSTO of NGC
1846 may be partly from the mass-accreted or merged binary stars.

\acknowledgments We thank the anonymous referee for his/her helpful
comments and acknowledge support from the CPSF 20100480222, NSFC
11003003, 10773003, 10933002 and the Ministry of Science and
Technology of the People's republic of China through grant
2007CB815406.

\clearpage

\clearpage

\clearpage


\begin{thebibliography}{}
\bibitem[Bastian \& de Mink(2009)]{bast09}
Bastian, N., \& de Mink, S. E. 2009, MNRAS, 398, L11
\bibitem[Bekki \& Mackey(2009)]{bekk09}
Bekki, K., \& Mackey, A. D. 2009, MNRAS, 394, 124
\bibitem[Bertelli et al.(2003)]{bert03}
Bertelli, G., Nasi, E., Girardi, L., Chiosi, C., Zoccali, M., \&
Gallart, C. 2003, AJ, 125, 770
\bibitem[Chabrier(2001)]{chab01}
Chabrier, G. 2001, ApJ, 554, 1274
\bibitem[D'Ercole et al.(2008)]{derc08}
D'Ercole, A., Vesperini, E., D'Antona, F., McMillan, S. L. W., \&
Recchi, S. 2008, MNRAS, 391, 825
\bibitem[Elson et al.(1998)]{elson98}
Elson, R. A. W., Sigurdsson, S., Davies, M., Hurley, J., \& Gilmore,
G. 1998, MNRAS, 300, 857
\bibitem[Girardi et al.(2000)]{gira00}
Girardi, L., Mermilliod, J. C., \& Carraro, G. 2000, A\&A, 354, 892
\bibitem[Girardi et al.(2009)]{gira09}
Girardi, L., Rubele, S., \& Kerber, L. 2009, MNRAS, 394, L74
\bibitem[Girardi et al.(2011)]{gira11}
Girardi, L., Eggenberger, P., \& Miglio, A. 2011, arXiv:1101.1880
\bibitem[Glatt et al.(2008)]{glat08}
Glatt, K. et al. 2008, AJ, 136, 1703
\bibitem[Goudfrooij et al.(2009)]{goud09}
Goudfrooij, P., Puzia, T. H., Kozhurina-Platais, V., \& Chandar, R.
2009, AJ, 137, 4988
\bibitem[Han et al.(1995)]{han95}
Han Z., Podsiadlowski P., \& Eggleton P.P. 1995, MNRAS, 272, 800
\bibitem[Hurley et al. (2000)]{hurl00}
Hurley, J. R., Pols, O. R., \& Tout, C. A., 2000, MNRAS, 315, 543
\bibitem[Hurley et al. (2002)]{hurl02}
Hurley, J. R., Tout, C. A., \& Pols, O. R. 2002, MNRAS, 329, 897
\bibitem[Keller et al. (2011)]{kell11}
Keller, S. C., Mackey, A. D., \& Da Costa, G. S. 2011,
arXiv:1102.1723

\bibitem[Lejeune et al. (1998)]{leje98} Lejeune, T.,
Cuisinier, F., \& Buser, R., 1998, A\&A, 130, 65
\bibitem[Mackey \& Broby Nielsen(2007)]{mack07}
Mackey, A. D., \& Broby Nielsen, P. 2007, MNRAS, 379, 151
\bibitem[Mackey et al.(2008)]{mack08}
Mackey, A. D., Broby Nielsen, P., Ferguson, A. M. N., \& Richardson,
J. C. 2008, ApJ, 681, L17
\bibitem[Milone et al.(2009)]{milo09}
Milone, A. P., Bedin, L. R., Piotto, G., \& Anderson, J. 2009, A\&A,
497, 755
\bibitem[Milone et al.(2010)]{milo10}
Milone, A. P. et al. 2010, ApJ, 709, 1183

\bibitem[Mucciarelli et al.(2008)]{mucc08}
Mucciarelli, A., Carretta, E., Origlia, L., \& Ferraro, F. R. 2008,
AJ, 136, 375
\bibitem[Piotto et al.(2005)]{piot05}
Piotto, G. et al. 2005, ApJ, 621, 777
\bibitem[Piotto et al.(2007)]{piot07}
Piotto, G. et al. 2007, ApJ, 661, L53
\bibitem[Rubele et al.(2010)]{rube10}
Rubele, S., Kerber, L., \& Girardi, L. 2010, MNRAS, 403, 1156
\bibitem[Rubele et al.(2011)]{rube11}
Rubele, S., Girardi, L., Kozhurina-Platais, V., Goudfrooij, P., \&
Kerber, L. 2011, arXiv:1102.2814
\bibitem[Salpeter(1955)]{salp55}
Salpeter E. E. 1955, ApJ, 121, 161
\end{thebibliography}
\end{document}